\documentclass[usenatbib]{mn2e}
\usepackage{graphicx}
\usepackage{aas_macros}
\usepackage{amsfonts}
\usepackage{bm}  %allows bold fonts for Greek letters
\usepackage{times}
\begin{document}

%%%%%%%%%%%%%%%%%%%%%%%%%%%%%%%%%%%%%%%%%%%%%%%%%%%%%%%%%%%%%%%%%%%%%%%%%
%%% Extra definitions 

\def\ns{\normalsize}
\def\nn{\nonumber}
\def\beqn{\begin{equation}}
\def\eeqn{\end{equation}}
\def\beqnarray{\begin{eqnarray}}
\def\eeqnarray{\end{eqnarray}}
\def\eps{\epsilon}
\def\LMC{\rm{LMC}}
\def\rsun{R_\circ}
\def\normn{{\bf{\hat{n}}}}
\def\norml{{\bf{\hat{l}}}}
\def\norms{{\bf{\hat{s}}}}
\def\normr{{\bf{\hat{r}}}}
\def\normrsun{{\bf{\hat{R}}_\circ}}
\def\vecrsun{{\bf{R}_\circ}}
\def\vecr{{\bf{r}}}
\def\vecl{{\bf{l}}}
\def\vecs{{\bf{s}}}
\def\vecd{{\bf{d}}}
\def\normsigma{{\bm{\hat{\sigma}}}}
\def\deg{\hbox{$\null^\circ$}}

\def\Mpc{\,{\rm Mpc}}
\def\kpc{\,{\rm kpc}}
\def\pc{\,{\rm pc}}
\def\Msun{\,{\rm M_\odot}}
\def\Gyr{\,{\rm Gyr}}
\def\Myr{\,{\rm Myr}}
\def\kms{\,{\rm km}\,{\rm s}^{-1}}

\def\Z{{\it{Z} }}

\def\ud{\mathrm{d}}
%%%%%%%%%%%%%%%%%%%%%%%%%%%%%%%%%%%%%%%%%%%%%%%%%%%%%%%%%%%%%%%%%%%%%%%%%

\title[Geometrodynamical Distances to the Galaxy's Hydrogen Streams]
{Geometrodynamical Distances to the Galaxy's Hydrogen Streams}

\author[S.~Jin \& D.~Lynden-Bell]{Shoko Jin\thanks{e-mail: shoko@ast.cam.ac.uk} \& D. Lynden-Bell\\
Institute of Astronomy, University of Cambridge, Madingley Road, Cambridge, CB3 0HA, U.~K.\\}

\maketitle

\begin{abstract}

We present a geometrodynamical method for determining distances to
orbital streams of HI gas in the Galaxy.  The method makes use of our
offset from the Galactic centre and assumes that the gas comprising
the stream nearly follows a planar orbit about the Galactic centre.
We apply this technique to the Magellanic Stream and determine the
distances to all points along it; a consistency check shows that the
angular momentum is approximately constant. Applying this technique to
the Large Magellanic Cloud itself gives an independent distance which
agrees within its accuracy of around 10\%.  Relaxing the demand for
exact conservation of energy and angular momentum at all points along
the stream allows for an increase in orbital period between the
lagging end and the front end led by the Magellanic Clouds.  Similar
methods are applicable to other long streams of high-velocity clouds,
provided they also nearly follow planar orbits; these would allow
otherwise unknown distances to be determined.

\end{abstract}


\section{Introduction}
\label{sec:intro}

High-velocity clouds (HVCs) are defined as HI gas which have anomalous
velocities that are incompatible with what is expected from a simple
Galactic rotation model.  Ever since their discovery by
\cite{HVCdiscovery}, their role in the formation and evolution of our
Galaxy has been a source of speculation.  Amongst the attempts to
explain their origin, the two most well-known and contested are the
Galactic Fountain model, in which gas is either being thrown out of
the Galactic disk from supernovae or being heated and ionized by
massive stars, and the alternative model in which the clouds are
pockets of neutral hydrogen left over from the formation epoch of the
Galaxy, and which are only now being accreted.

Given the vast sky-coverage of HVCs, the two theories are not
necessarily contradictory, as it is likely that some clouds
belong to each model.  However, with the exception of those HVCs 
that are obvious components of a stream with a known progenitor, such
as the Magellanic \citep{1974ApJ...190..291M} and Sagittarius streams
\citep{2004ApJ...603L..77P}, the lack of distance measurements has
hindered the progress in understanding the role that HVCs play in the
Galaxy.

Since newly accreted gas would have a lower metallicity than recycled
gas from the Galaxy, metallicity measurements, if available, should be
able to distinguish the applicability of each theory on a
cloud-by-cloud basis
\citep[e.g.][]{2002ApJ...572..178S,2000ASPC..218..407V,2001ApJS..136..463W},
but it would still be instructive to know exactly where these clouds
lie.  Observationally, the difficulty in determining direct distances
arises from the fact that most of the HVCs do not host stars.  If,
however, there is one star behind and another to the front of the
cloud along the line of sight, both of which have known distances,
then a distance bracket can be set for the HVC.  The high resolution
spectrum of a star directly behind the HVC will exhibit absorption
lines at the velocity of that cloud; features that will be absent in
the spectrum of a star to the front of the cloud.  Starting with the
first such measurement for Complex M in the early 90's
\citep{1993ApJ...416L..29D}, other HVC complexes for which a distance
bracket that now exists include Complex A \citep{2001ApJS..136..463W}
and Complex WB \citep{2006ApJ...638L..97T}.  The high latitudes of
much of the neutral hydrogen gas in question means there is a lack of
background stars bright enough for high resolution spectroscopy, so
this technique has not been without difficulty.

Many detailed simulations of streams such as the Magellanic Stream
exist
\citep[e.g.][]{2006MNRAS.371..108C,2005MNRAS.363..509M,1996MNRAS.278..191G,
  1985IAUS..106..471F,1976A&A....47..263F}, however, all such
simulations have many parameters which must be tweaked to give the
best fits to the data.  In this paper, we describe an alternative set
of methods with the aim of probing the elusive distances to streams of
HI gas comprising the HVCs.  These methods instead make strong
hypotheses that enable us to attempt to read the distances out of the
data themselves; the studies presented here are neither simulations
nor attempts at modelling.  The following section introduces the
geometrodynamical methods for determining distances to streams of HVCs
and discusses which streams are appropriate for each method.  We
developed the methods in response to a challenge laid down by
Majewski, who asked whether it was possible to determine distances
from radial velocities alone, when those were known along a stream
\citep{2004ASPC..327...63M}.  In Section \ref{subsec:masterpoints}, we
consider distant streams and use heliocentric radial velocities
corrected to the Galactic Standard of Rest (GSR) as a lowest order
approximation to the Galactocentric radial velocities. Section
\ref{sec:velcor} shows how these radial velocities can be corrected
iteratively, and shows that the corrections are negligible for the
Magellanic Stream.  The choice of Galactic potential is briefly
discussed in Section \ref{sec:potential}.  On the assumption that the
Magellanic Stream is composed of material that was torn off the
Magellanic Clouds at an earlier passage through perigalacticon, it
seems reasonable to assume an angular momentum gradient to be present
along the length of the stream; this topic is addressed in Section
\ref{sec:deltah}.  We present our summary and conclusions in Section
\ref{sec:sum}.

There is a great body of literature on streams and associations of
bodies in planes through the Galactic centre.  The Sagittarius stream
\citep{2001ApJ...547L.133I,2003AJ....125.1352P} features prominently
in this literature, however it is not relevant to the methods
developed here which work off the parallax due to the Sun's
displacement from the plane of the orbit.  This displacement is
almost zero for the Sagittarius stream.

\begin{figure}
\hspace{-10pt}
\includegraphics[width=0.5\textwidth]{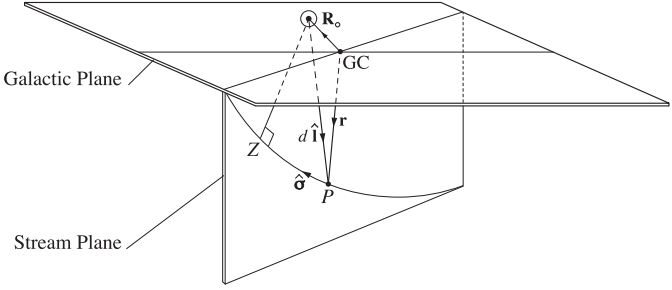}
\caption{An orbital stream lying in a plane
 through the Galactic centre.  The unit vector ${\rm {\bm{\hat{\sigma}}}}$
 lies along the stream at a general point $P$, which is seen from the
 Sun in the direction ${\bf{\hat{l}}}$ at a distance $d$.  Its observed
 radial velocity, corrected to the Galactic Standard of Rest, is $v_l$ =
 ${\bf{v}}.{\bm{\hat{l}}}$; at the position labelled $Z$,
 $v_l$ = 0. 
The unit vector $\norms$ is along the projection of $\normsigma$ onto
the sky, so $\norms.\norml = 0$ and $\norms$ is $\normsigma -
(\normsigma.\norml)\norml$ normalised to unity.  At $Z$, $\norms$ and
$\normsigma$ coincide.}
\label{fig:schematic}
\end{figure}

Some frequently used symbols and their definitions are summarised in
Table \ref{tab:defs} with a schematic guide to a generic point on the
stream given by Figure \ref{fig:schematic}.

\begin{table*}
\caption{ Definition of commonly used symbols, where hats denote unit vectors. \label{tab:defs} }
\begin{tabular}{ccl}
\hline
  Symbol
& \multicolumn{1}{c}{Equation where first used}
& \multicolumn{1}{c}{Description} \\
\hline
$\norms$, $\normsigma$ & (\ref{eqn:theta}),  (\ref{eqn:v}) & Apparent and true direction vectors for a particular point on the stream\\
$\normn$ & (\ref{eqn:theta}) & Vector normal to the stream plane\\
$\norml$, $\normr$ & (\ref{eqn:theta}), (\ref{eqn:d}) & Unit vectors to a stream point, from the Sun and the  Galactic centre\\
$v_l$ & (\ref{eqn:v}) & Heliocentric line-of-sight velocity corrected to the Galactic System of Rest\\
$\vecrsun$ & (\ref{eqn:d}) & Vector joining Galactic centre to Sun $=(-\rsun,0,0)$ \\
$d$, $r$ & (\ref{eqn:d}), (\ref{eqn:energy}) & Distance to a specific point on the stream from the Sun and from the Galactic centre\\
${\bf{h}}$, $\varepsilon$ & (\ref{eqn:am}), (\ref{eqn:energy}) & Specific angular momentum and energy associated with the stream \\
$D$ & (\ref{eqn:norm}) & Ratio of distances to the $Z$ point and the Galactic centre from the Sun \\
$\psi$ & (\ref{eqn:energy}) & Galactic potential \\
$v_c$ & (\ref{eqn:rH}) & Galactic circular velocity at $\rsun$ (unless specified otherwise)\\	
\hline
\end{tabular}
\end{table*}

\section{Distance Determination Methods}
\label{sec:methods}

In the following sections, we explore several different variations of
our geometrodynamical distance-determination scheme and their
applicability.

\subsection{Near Method}
\label{subsec:near}

Let us suppose that we have a set of observed heliocentric radial
velocities relative to the Local Standard of Rest (LSR) along a
neutral hydrogen stream, which can be re-expressed in the GSR.  We
define a location $Z$ to be a point on any stream where the
line-of-sight velocity as seen in the GSR, $v_l$, is zero.
The assumptions made are that the clouds of neutral hydrogen form a
single stream, the streaming material is in a plane containing the
Galactic centre, and that the stream is moving along itself.  If these
assumptions are valid, then the implication is that the
line-of-sight vector to $Z$ lies perpendicular to the true
direction of motion of the stream there.  Here, the apparent
direction will equal the true direction of the stream's motion.  Hence
for a given $Z$ point at any assumed distance of $d_Z$ from the Sun, we
will have two vectors that lie in the plane of the stream: the
vector from the Galactic centre to $Z$ and the directional vector,
${\bf{\hat{s}}}_Z$, in the plane of the sky.  If $d_Z$ can be found,
the plane of the entire orbit could be found from the cross-product of
these two vectors.  All other points along the stream could then be
projected along the lines of sight onto this plane, enabling their 
distances and all three components of their velocities to be found.

The distance to the $Z$ point may be found by looking for a value
which fulfills the requirements of energy and angular momentum
conservation along the stream, assuming that such a value exists.

Let us consider a point on the stream, to which we have a
line-of-sight vector $\vecl$\footnote{The vectors, unless specified
otherwise, are stated in the Galactic coordinates, whose origin is
defined as the location of the Sun, and whose $x$ direction lies
towards the Galactic centre.  The $z$ axis points to the North
Galactic Pole.}.  We denote the unit vector describing the apparent
direction of the stream on the sky as $\norms$.  Here and hereafter,
hats denote unit vectors.  If $\normsigma$ is defined to be the true
direction of the stream at this location and $\normn$ the normal to
the stream plane, then by construction, $\norml.\norms = 0$ and
$\normsigma = \norms\cos\theta + \norml\sin\theta$, where $\theta$ is
the angle between the apparent direction and the true direction.
Clearly, $\normsigma$ lies in the stream's orbital plane, and hence
$\theta$ satisfies

\beqn
\label{eqn:theta}
\cot\theta = -\frac{\norml.\normn}{\norms.\normn}\,.
\eeqn
The velocity of the stream is, by hypothesis, parallel to
$\normsigma$, and hence may be written as ${\bf{v}} = v\normsigma$.  The
line-of-sight velocity in the GSR is then given by $v_l =
{\bf{v}}.\norml = v\sin\theta$.  This gives for
the velocity:

\beqn
\label{eqn:v}
{\bf{v}} = v_l\,\frac{\normn\times(\norml\times\norms)}{\norms.\normn}
\hspace{0.8cm}{\rm for}\hspace{0.8cm}
\norml.\normsigma\neq0\,.
\eeqn   
At the $Z$ point, both $v_l$ and $\theta$ are zero.  Let us denote
quantities associated with this point with a subscript $Z$ and let
$\vecrsun$ be the vector from the Galactic centre to the Sun.  If
$\vecr$ is the vector joining the Galactic centre to a general point
in the stream at distance $d$ from the Sun, then $\vecr = \vecrsun +
d\,\norml$ and, specifically at $Z$, $\vecr_Z = \vecrsun +
d_Z\,\norml_Z$. By definition, $\vecr$ lies in the stream's orbital plane,
and hence:

\beqn
\label{eqn:d}
0 = \vecr.\normn = \vecrsun.\normn + d\,\norml.\normn
\hspace{0.8cm}\Rightarrow\hspace{0.8cm}
d = -\frac{\vecrsun.\normn}{\norml.\normn}\,,
\eeqn
and we can eliminate $d$ to obtain an expression for $\vecr$ in terms
of only $\vecrsun$, $\norml$ and $\normn$:
\beqn
\label{eqn:r}
\vecr = \vecrsun + d\,\norml = \frac{\vecrsun(\norml.\normn) - \norml(\vecrsun.\normn)}{\norml.\normn} = \frac{(\norml\times\vecrsun)\times\normn}{\norml.\normn}\,.
\eeqn

Equipped with these vector relations, we now show how conserving the
angular momentum along the stream enables us to find the distance to
the $Z$ point, which serves as a representative distance to a stream.

The specific angular momentum (henceforth referred to as the `angular
momentum') associated with the orbit of a stream is given by 
\beqnarray
\label{eqn:am}
{\bf{h}} &=& {\bf{r}}\times{\bf{v}} \nn\\
 &=& [(\norml\times \normrsun)\times\normn]\times[\normn\times(\norml\times\norms)]\,\rsun\,v_l\, 
/\,[(\norml.\normn)(\norms.\normn)]\nn\\   
 &=& [\normrsun.(\norml\times\norms)/(\norms.\normn)]\,\rsun\,v_l\,\normn\,.
\eeqnarray
Conservation of angular momentum implies that the coefficient of
$\normn$ is constant along the stream.  Since $\normrsun$ and $v_l$
are known, the only remaining unknown is $\normn$, assuming that $\norms$
is known at each point of the stream from looking at the sky.  Since
the normal to the plane is parallel to ${{\bf{r}}_Z}\times\norms_Z =
(\vecrsun + d_Z \norml_Z)\times\norms_Z$, it may be written:

\beqn
\label{eqn:norm}
\normn = (\vecr_Z\times\norms_Z)/|\vecr_Z\times\norms_Z| 
= \rsun\frac{[\normrsun\times\norms_Z + D\,\norml_Z\times\norms_Z]}{|\vecr_Z\times\norms_Z|}\,,
\eeqn
where $D = d_Z/\rsun$ is unknown.  On the basis that ${\bf{h}}$ is
constant, we can derive an expression to determine $D$ through other
observable or calculable quantities.  Since ${\bf{h}}$ is parallel to
$\normn$, ${\bf{h}}.\normn$ is a constant.  Taking the dot product of
${\bf{h}}$ in equation (\ref{eqn:am}) with $\normn$ gives an
expression in which $\norms.\normn$ can then be expanded via $\normn$
given in equation (\ref{eqn:norm}).  This resulting expression is then
inverted to give the following condition along a planar stream:

\beqn
\label{eqn:line}
\frac{[\normrsun\times\norms_Z + D\,\norml_Z\times\norms_Z] \,.\, \norms}{\normrsun . (\norml\times\norms)\,\rsun\,v_l} = \rm{constant}\,\,.
\eeqn
Setting
\beqn
\label{eqn:XY}
Y = \frac{(\normrsun\times\norms_Z)\,.\,\norms}{\normrsun . (\norml\times\norms)\,\rsun\,v_l}
\hspace{0.4cm}{\rm{and}}\hspace{0.4cm}
X = \frac{(\norml_Z\times\norms_Z)\,.\,\norms}{\normrsun . (\norml\times\norms)\,{\it \rsun\,v_l}}
\,,
\eeqn
plotting $Y$ against $X$ should give a straight line of slope $-D$.

\begin{figure}
\hspace{-10pt}
\includegraphics[width=0.5\textwidth]{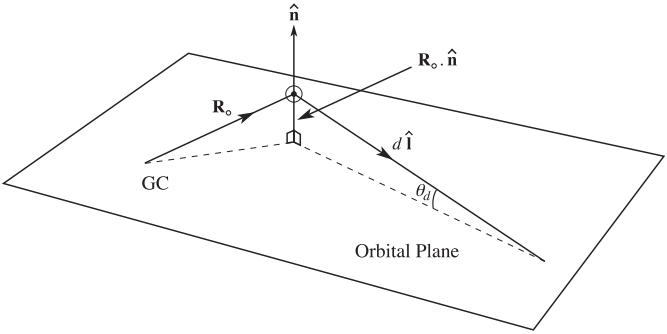}
\caption{A schematic diagram showing the angle $\vartheta_d$ subtended
  at the stream by $\vecrsun$ projected onto $\normn$, the normal to
  the stream's orbital plane.  $\vecrsun$ joins the Sun to
  the Galactic centre.  $\vartheta_d$ is given by
  $\mathrm{arcsin}(\vecrsun.\normn/d)$ and is useful in
  quantifying the range of validity of the near method -- see main
  text for details.}
\label{fig:theta_d}
\end{figure}

The offset of the Sun from the Galactic centre produces insufficient
parallax for this method to work if the stream plane passes close to
$\vecrsun$.  We can see this again from equation (\ref{eqn:XY}), as
both the numerators and denominators vanish when $\vecrsun$ lies in
the stream plane.  We performed tests on simulated orbits of varying
orientations, which were given some random scatter in direction
cosines in order to test the accuracy to which $\norms$ was required.
We found that the angle $\vartheta_d$ subtended at the stream by
$\vecrsun$ projected onto $\normn$ needed to be at least $9\deg$ if
errors in the determination of $\norms$ and $\norms_Z$ are not to
dominate in equation (\ref{eqn:line}).  Figure \ref{fig:theta_d} shows
how $\vartheta_d = \mathrm{arcsin}(\vecrsun.\normn/d)$ is subtended at
the stream.  It is also true that if $D$ is very large, then errors in
$X$ may dominate in determining the gradient of the line $Y + D\,X =
\rm{const}$.  Only lower limits to $D$ could then be determined.  In
general, an `ideal' data set will give the expected straight line,
with singular behaviour in the form of another line (with positive or
negative gradient) for points which are close to $Z$.  This enables
$d_Z$ to be determined reliably to an accuracy of within a few per
cent.

In practice, the data for the Magellanic Stream did not produce a
straight line.  With hindsight we might have expected this, since the
method should not work for a distant stream for which we have a very
radial view and very little direct information of the transverse
velocity components.  The radial velocities do not contain enough
information on the angular momentum of the orbit if the stream is
distant and, for most parts of the stream, $\vartheta_d$ subtended at the
stream becomes too small for $\norms$ to be determined reliably.
Ironically, this angle is largest close to $Z$, where the denominator
in equation (\ref{eqn:line}) vanishes!  The method should be better
suited to closer streams but there the assumption of a planar orbit is
more doubtful unless the orbit passes close to the Galactic Pole.  We
note in passing that the plane of the Sagittarius Stream lies much too
close to $\vecrsun$ for it to be a suitable candidate.

Our first attempt to use this method on Complex A produced rather
perplexing results.  We found a good straight line between the $X$ and
$Y$ variables of equation (\ref{eqn:XY}), but with a positive slope,
rather than the negative slope that we would expect.  The solution
gives a distance of $-5\kpc$, in other words in the opposite direction
to which we see it.  Clearly this was nonsense.  Having tested this
near method on simulated orbits, we realised there were two possible
explanations as to why the method might not work for Complex A, even
if the stream was following an orbit.  The angles $\vartheta_d$ at all
points along the stream are large enough at distances that we might
expect to find for Complex A.  However, there is a competing effect
that these points are all rather close to the $Z$ point, and the
denominators in equation (\ref{eqn:XY}) vanish at $Z$.  On the other
hand, it is possible that the stream is simply too close to be in a
well-defined plane.  The flattening of the potential due to the disk
would be a significant deviation from the assumption of a spherical
potential.

We recently posed the question of whether the stellar Orphan Stream
\citep{2007ApJ...658..337B} and Complex A might be related
\citep{2007MNRAS.378L..64J}.  By applying the near method presented
here to the orbit for Complex A found in this work, we found that the
perplexing positive slope could be explained, in that case, as part of
the singular behaviour that is expected with points close to $Z$.
We also tried the near method on Complex C, which has two $Z$ points;
see Section \ref{subsec:2Z} for an interpretation of the results.

In summary, the range of validity of this first method of distance
determination is to systems with orbital planes well away from the
Sun-Galactic centre line but for which the angle $\vartheta_d =
\mathrm{arcsin}(\vecrsun.\normn/d)$ subtended at the stream points is
at least $9\deg$.  This leads to distances not more than a factor of
order six larger than $\rsun$ for a stream of Magellanic Stream-like
orientation.  In Section \ref{sec:nearmethodex}, we give a working
example of the near method.

The advantage in attempting to use conservation of just the component
of angular momentum about the Galactic axis is that it requires no
detailed model of the Galactic potential in which the stream flows.
We have, however, identified some conditions under which the near
method via angular momentum conservation does not work.  We considered
next the implications of energy conservation, which should not have
suffered from the same issues.  On applying this method to Complex~A,
we found that the $\norms$ vectors were too inaccurately determined to
find a planar orbit which gave a consistent distance to the $Z$ point,
partly because Complex A extends only $50\deg$ across the sky with the
$Z$ point to one end.  Points near $Z$ give inaccurate results as
they suffer from small denominators.

\subsection{Far Method}
\label{subsec:masterpoints}

Application of the method described in Section \ref{subsec:near} to
the Magellanic Stream caused some problems.  We need to determine the
plane of a stream to such an accuracy that the line of sight to the
clouds can be projected to hit that plane at their true distances.
Lines of sight at small angles to the plane lead to inaccurate
distances due to small offsets of the clouds from the best-fitting
plane.  The next method circumvents this by using the line-of-sight
vectors to individual points on the stream.

When the points of a stream are at distances from the Sun much greater
than $\rsun$, then to first approximation the heliocentric
line-of-sight radial velocities corrected to the GSR, $v_l$, will be
close to $\dot{r}$, the Galactocentric\footnote{The Galactocentric
coordinate system is defined by translating the Galactic coordinate
system along its $x$ axis by $\rsun$ so that the coordinate origin
coincides with the Galactic centre.} radial velocities.  On the
hypothesis that the clouds of the stream follow an orbit in a plane
through the Galactic centre, the energy equation in a potential
$\psi({\bf{r}})$ is given by

\beqn
\label{eqn:energy}
\frac{{\bf{v}}.{\bf{v}}}{2}  - \psi({\bf{r}}) = \varepsilon\,,\\
\eeqn
where $\varepsilon$ is the specific energy (hereafter referred to as
the `energy').  If there is a cloud at which the stream reaches a
maximum value of $|v_l|\simeq|\dot{r}|$, then labelling such a point
the $H$ point and differentiating equation (\ref{eqn:energy}), we find
$h^2/r_H^3 = -\partial\psi/\partial r |_{r_H} \equiv v_c^2(r_H)/r_H$,
giving $h = r_H v_c(r_H)$.  This relationship is particularly simple
for the special case when $v_c = {\rm constant}$, in which case a
simple logarithmic potential $\psi(r) = -v_c^2\ln r/r_h$ can be used,
for some halo cut-off radius $r_h$; we will therefore work with that
special case first and return to the more general case in Section
\ref{sec:potential}.  Subtracting the energy equation at a general
point from that at $H$, we have, using the above value for $h$ and
dividing by $v_c^2/2$:

\beqn
\label{eqn:rH}
\frac{\dot{r}_H^2 - \dot{r}^2}{v_c^2} + 1 = \frac{r_H^2}{r^2} - \ln\frac{r_H^2}{r^2}\,,
\eeqn
where $v_{l,H}$ and $v_l$ are used for $\dot{r}_H$ and $\dot{r}$
respectively and $v_c$ is set to $220\kms$.  We may then solve this
equation for $r/r_H$.  Note that there are two possible roots to
the equation, one on each side of $r_H$, but in practice $r<r_H$ in
the Magellanic Stream, which eliminates one solution.  This would
clearly change if clouds existed past the $H$ point.

\begin{figure}
\includegraphics[width=0.45\textwidth]{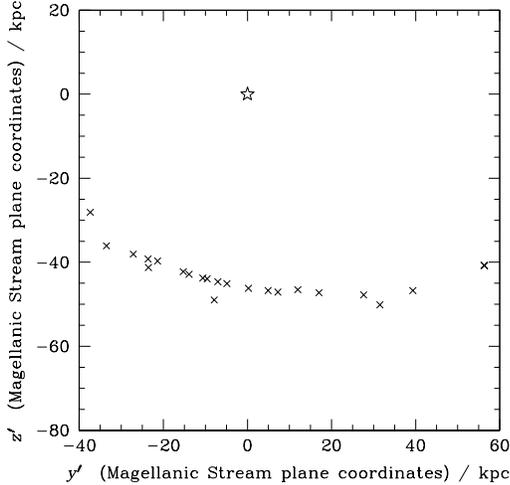}
\caption{The Magellanic Stream depicted in its own plane, with the
leading end (position of the Magellanic Clouds) to the left of the
plot.  The Galactic centre is denoted by the star.  The
$y\prime$ and $z\prime$ axes shown here are nearly parallel to the
Galactic $y$ and $z$ axes respectively.  There is little scatter in
the third (unshown) dimension, the axis for which lies along the
normal to the plane.
\label{fig:stream}}
\end{figure}

\begin{figure}
\includegraphics[width=0.45\textwidth]{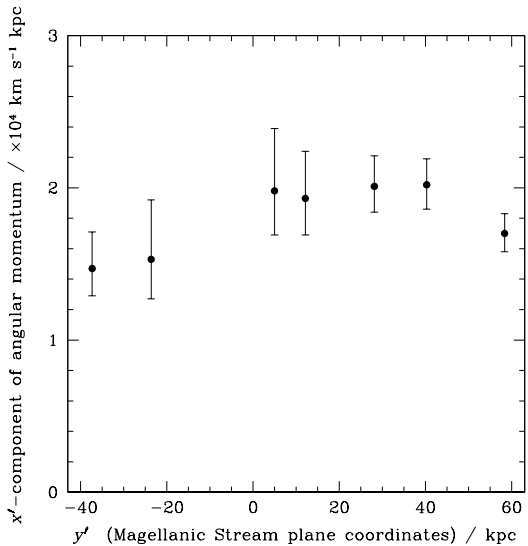}
\caption{The magnitude of the $x\prime$-component of the angular
momentum for points along the Magellanic Stream is plotted against
distance along the stream in the $y\prime$ direction.  An explanation
of the error bars is given in the main text.
\label{fig:angmom}}
\end{figure}

Equation (\ref{eqn:rH}) gives ratios of distances from the Galactic
centre to points on the stream.  The distances from us, $d$, are then
given in terms of $r_H$ by
\beqnarray
\label{eqn:d3}
d^2 - 2d\norml.\vecrsun + \rsun^2 = \left(\frac{r}{r_H}\right)^2 r_H^2\,\,,
\eeqnarray
so for three points $r_i$ on the stream, $i$=1, 2 and 3,
\beqnarray
d_i = \norml_i.\vecrsun + \left(\left(\frac{r_i}{r_H}\right)^2\left(\frac{r_H}{\rsun}\right)^2 - \rsun^2 + (\norml_i.\vecrsun)^2\right)^{1/2}\,\,.
\eeqnarray
The condition that three points with distance vectors ${\bf {d}}_i =
d_i\norml_i$ lie in a plane through the Galactic centre is that for these
points, $\vecd_i + \vecrsun$ lies in the plane of the stream.  Hence we have

\beqn
\left[(\vecd_1 + \vecrsun)\times(\vecd_2 + \vecrsun)\right].(\vecd_3 + \vecrsun) = 0\,.
\eeqn
Rearranging this, we can then solve for the value of $r_H/\rsun$ that
satisfies

\beqn 
\frac{\vecrsun.\left(\vecd_1\times\vecd_2 + \vecd_2\times\vecd_3
+ \vecd_3\times\vecd_1 \right)}{(\vecd_1\times\vecd_2).\vecd_3} = -
1\,.\\ 
\eeqn 
Using equation (\ref{eqn:rH}), distances may be found to the rest of
the stream clouds, for which velocity data are available.  The mean
line through the observed velocities as a function of angle along the
primary stream direction reaches $205\kms$ some $74\deg$ behind the
$Z$ point, which we take to be the $H$ point since it has the highest
velocity of any substantial hydrogen on that mean line.  For $\rsun =
8.5\kpc$, a typical set of values found is $r_H = 75\kpc$ and $\normn
= (0.995, 0.080, -0.066)$.  The choice of the three fiducial points is
somewhat arbitrary, but to get good angular leverage, we take two
points near the ends of the stream and one toward the centre.  Our
results for the Magellanic Stream are given in Figures
\ref{fig:stream} and \ref{fig:angmom} and Table \ref{tab:stream}.  We
find that the perigalacticon is located at approximately $45\kpc$ and
our last point on the stream (just prior to the $H$ point) is $70\kpc$
from the Galactic centre.  The centre of mass of the Magellanic Clouds
is independently determined to be at $47\pm 5\kpc$.  The error in
determining the distances is approximated by the ratio of the
uncertainty in the central line of the stream to the parallax due to
the Sun's offset from the Galactic centre, which is about $9\%$.

\begin{table}
\caption{Galactocentric distances to the Magellanic Stream.\label{tab:stream}}
\begin{tabular}{rrrrrrrr}
\hline
 \multicolumn{1}{c}{$l /\deg$}
& \multicolumn{1}{c}{$b /\deg$}
& \multicolumn{1}{c}{$v_l /$}
& \multicolumn{1}{c}{$r + \delta r /$}
& \multicolumn{1}{c}{$x /\kpc$}
& \multicolumn{1}{c}{$y /\kpc$}
& \multicolumn{1}{c}{$z /\kpc$}\\
  \multicolumn{1}{c}{}
& \multicolumn{1}{c}{}
& \multicolumn{1}{c}{$\kms$}
& \multicolumn{1}{c}{$\kpc$}
& \multicolumn{1}{c}{}
& \multicolumn{1}{c}{}
& \multicolumn{1}{c}{}\\

\hline

  284	&-36    &83	&46.8 $-$ 0.1 &  0.8  &-37.4  &-28.0 \\    
\dag\,285&-46   &122	&49.3 $-$ 0.2 &  0.5  &-33.6  &-36.0 \\ 
296	&-52    &88	&47.0 $-$ 0.0 &  4.6  &-26.9  &-38.3 \\    
  292	&-57    &59	&45.8 $+$ 0.0 &  1.0  &-23.6  &-39.2 \\    
  295	&-58    &99	&47.6 $-$ 0.1 &  2.4  &-23.4  &-41.4 \\    
  290	&-60    &23	&45.1 $+$ 0.1 &  -0.7 &-21.5  &-39.6 \\   
  302	&-67    &7	&45.0 $+$ 0.2 &  1.0  &-15.2  &-42.3 \\   
  295	&-70    &-18	&45.0 $+$ 0.2 &  -1.9 &-14.1  &-42.7 \\   	
  300	&-74    &-15	&45.0 $+$ 0.2 &  -2.2 &-10.8  &-43.6 \\   
  315	&-73    &-20	&45.1 $+$ 0.2 &  1.0  & -9.5  &-44.0 \\   	
  310	&-78    &-30	&45.2 $+$ 0.3 &  -2.4 & -7.3  &-44.5 \\   
  300	&-79    &-126	&49.6 $+$ 0.8 &  -3.8 & -8.2  &-48.8 \\   
  320	&-80    &-40	&45.3 $+$ 0.3 &  -2.4 & -5.1  &-45.0 \\   
\dag\,\,\,0&-83 &-70	&46.2 $+$ 0.4 &  -2.8 &  0.0  &-46.1 \\   
   45	&-82    &-88	&47.0 $+$ 0.5 &  -3.9 &  4.6  &-46.6 \\   
   57	&-80    &-100	&47.6 $+$ 0.5 &  -4.0 &  6.9  &-47.0 \\   
   70	&-75    &-106	&48.1 $+$ 0.5 &  -4.2 & 11.7  &-46.4 \\   
   77	&-70    &-133	&50.2 $+$ 0.7 &  -4.6 & 16.7  &-47.1 \\   
\dag\,\,83&-60  &-168	&55.2 $+$ 1.0 &  -5.1 & 27.3  &-47.7 \\   
   85	&-58    &-184	&59.2 $+$ 1.3 &  -5.8 & 31.1  &-50.0 \\   
   85	&-50    &-189	&61.1 $+$ 1.5 &  -5.1 & 39.0  &-46.7 \\   
   87	&-36    &-202	&69.6 $+$ 3.2 &  -5.6 & 56.1  &-40.8 \\   	
   88	&-36    &-202	&69.4 $+$ 2.6 &  -6.5 & 55.9  &-40.7 \\   
\hline
\end{tabular}		     
\caption{Columns 1 and 2 give the Galactic longitude and latitude
respectively; column 3 gives the heliocentric line-of-sight velocities
corrected to the Galactic Standard of Rest; column 4 gives the
uncorrected distance $r$ from the Galactic centre in the plane
(i.e. $\sqrt{y\prime^2 + z\prime^2}$) and the correction $\delta r$
(see Section \ref{sec:velcor}); columns 5, 6 and 7 together give the
position vector in the Galactocentric coordinate system.  The data
coincide with the order of points in Figure \ref{fig:stream}, starting
from the left.  Note that $\delta r$ has been rounded to the nearest
decimal.  The fiducial points used to define the stream plane are
denoted with \dag.}
\end{table}

\subsection{Two $Z$ Points}
\label{subsec:2Z}

If a stream exhibits two $Z$ points, as suggested by the velocity field
of Complex C presented by \citet{2004hvc..conf...55S}, we have yet
another possible variation on a theme.

\subsubsection{Method}

Let us denote the unit directional vector for a $\Z_i$ point as
${\bf\hat{s}}_i$.  Then, by using the fact that ${\bf\hat{s}}_i$ for
both \Z points lie in the plane of the stream, we can find the normal
${\bf\hat{n}}$ to the plane as

\beqn
\label{eqn:norm2}
{\bf\hat{n}} = \frac{{\bf\hat{s}_1} \times {\bf\hat{s}_2}}{|{\bf\hat{s}_1} \times {\bf\hat{s}_2}|}\,.\\
\eeqn
As before, the absolute distance of the plane from us is found by
constraining the Galactic centre to be in the plane.  We know that at
the \Z points, ${\bf\hat{l}}_i.{\bf\hat{s}}_i = 0$, and equation
(\ref{eqn:d}) applies to each of the points.  The problem then reduces
down to finding the ${\bf \hat{s}}_i$ vectors.  From page 31 of
\cite{hvcbook2004}, we have an all-sky Aitoff projection map of the
high-velocity clouds.  By eye, it is possible to draw what would
appear to be the directional line of the stream, especially as the
stream is so long.  We can hence draw a directional `vector' from each
of the \Z points.  Consider the first \Z point given by Galactic
coordinates $(l_{Z1},b_{Z1})$ and a close-by point along this
`vector', whose coordinates are given by $(l_{P1},b_{P1})$; ${\bf
s}_1$ will then be given by the vector joining these two points.

Consider the celestial sphere of radius $r=1$.  Then ${\bf s}_1 =
R{\bf\hat{l}}_{P1} - r{\bf{\hat{l}}}_1$ where $R\neq 1$.  For the
moment, $R$ is an unknown.  However, since ${\bf\hat{l}}_1.{\bf s}_1 =
0$, then by substitution and rearrangement, we find that $R =
({\bf\hat{l}}_{1}.{\bf\hat{l}}_{P1})^{-1}$.  This implies that

\beqn
{\bf s}_1 = \frac{ {\bf\hat{l}}_{P1} }{ {\bf\hat{l}}_{P1}.{\bf\hat{l}}_1 } - {\bf\hat{l}}_1\,.\\
\eeqn
By evaluating the cross product ${\bf s}_1\times{\bf s}_2$ and
normalising, the unit normal to the plane may be found.

We now have a way of determining the distance to any point on the
stream by using equation (\ref{eqn:d}), which is general and not
confined to the \Z points.  It does not assume that the path of the
stream is confined to an orbit of constant energy and angular
momentum, however we require the stream to be in a planar
configuration as in all previous methods.

\subsubsection{Results}

\begin{figure}
\includegraphics[width=0.45\textwidth]{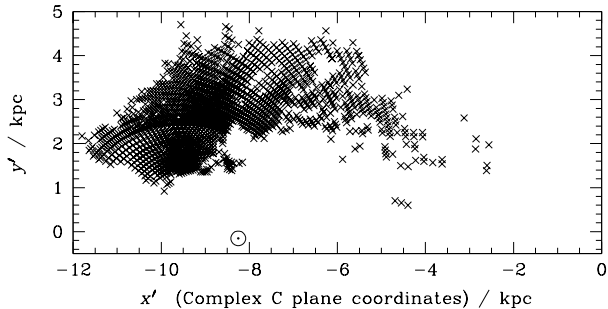}
\caption{Complex C depicted in its own plane.  The Galactic centre
corresponds to the coordinate location (0,0) and the position of the
Sun is projected onto the plane and shown by $\odot$.  The $x\prime$
and $y\prime$ axes shown here are nearly parallel to the Galactic $x$
and $y$ axes respectively.
\label{fig:complexc}}
\end{figure}

The velocity field of Complex C is given by Figure 4 of
\citet{2004hvc..conf...55S}, from which we interpolate the following
two \Z points: $(l_{Z1},b_{Z1}) = (37\deg, 18\deg)$ and
$(l_{Z2},b_{Z2}) = (110\deg, 48\deg)$.  With the following choice for
associated points to find the apparent direction vectors for each of
the \Z points --- $(l_{P1},b_{P1}) = (13\deg, 11\deg)$ and
$(l_{P2},b_{P2}) = (90\deg, 42\deg)$ --- we find the normal to the
supposed plane of this complex to be $(0.252,-0.071,0.965)$ and a set
of distances to the stream with a range of $\sim 3-6\kpc$ under this
method, as shown in Figure \ref{fig:complexc}.  The stream has a
significant width, across which there is some non-negligible scatter
in the velocities. 

It is slightly concerning that some of the clouds are rather close-by
and yet they already have negative line-of-sight velocities.  For
example, the range of GSR velocities at $x' = -5\kpc$ ranges from -20
to$ 20\kms$, while at $-4\kpc$ the range is -50 to $-20\kms$.  It is
somewhat unlikely that these clouds would be able to wrap around to
the other side of the Galactic centre with such infalling velocities.
The near method in Section \ref{subsec:near} applied to Complex C gave
similar but slightly smaller distances.  Both the current treatment
and the near method have likely given distances which are too small to
be plausible. We deduce that Complex C is not a stream lying in a
plane through the Galactic Centre.

\section{Correction of Radial Velocities}
\label{sec:velcor}

The method discussed in Section \ref{subsec:masterpoints} was
revisited, but with a correction now applied to the velocities in
order to account for our offset from the Galactic centre.  Whereas
previously the heliocentric line-of-sight velocities corrected to the
GSR were used in the energy equation, we now corrected these
iteratively to actual Galactocentric radial velocities.  We first
calculated, using line-of-sight velocities, the zeroth order distances
to the stream clouds.  With the three-dimensional configuration of the
stream then known to lowest order, we could transform to planar
coordinates with the suppressed axis corresponding to the normal to
the stream plane as defined by the three fiducial points (see Section
\ref{subsec:masterpoints} for definition).  We discounted the third
dimension --- vertical offsets from the stream plane --- on the basis
that these deviations were small relative to both components in the
plane and overall distance of the stream from the Galactic centre. We
then know the approximate direction in which a particular point in the
stream is moving, assuming no motion into or out of the plane.  This
assumption was reasonable given the size of the vertical offsets.  The
component of the true velocity projected along our line of sight gives
the observed velocity $v_l$, since radial velocities are taken to be
positive if they are outwards, hence an appropriate de-projection of
$v_l$ gives the `true' velocity.  This can then be re-projected along
the radial vector from the Galactic centre to the stream point to give
the corrected radial velocity.

We found the line-of-sight velocities to have been a slight
under-estimate on the Galactocentric radial velocities for the
Magellanic Stream.  The maximum change in Galactocentric distance as a
result of this velocity correction is approximately $3.2\kpc$ for the
farthest point, and the smaller distances are less affected, on
average by less than $1\kpc$.  We conclude that the size of the
corrections, which are presented in Table \ref{tab:stream}, are small
in comparison with errors that might result from the choice of method
to derive the distances.

By calculating the `true' velocity vector, we were also able to
calculate the angular momentum vector at each point along the stream.
In Figure \ref{fig:angmom}, the magnitude of the $x\prime$-component
of the angular momentum is plotted as a function of distance along the
stream, where the $x\prime$ axis is parallel to the normal to the
stream plane, and nearly parallel to the Galactic $x$ axis.

In order to determine the component of velocity along the stream, we
require the angle between the stream and the vector transverse to the
radial vector.  This can be determined from Figure \ref{fig:stream}.
Since the angular momentum depends on the cotangent of this angle, the
errors on the angular momenta are derived from assuming a $\pm2\deg$
error in this angle.  Near the perigalacticon, the angle becomes close
to $90\deg$.  For such points, the associated errors become very
large; these points have not been plotted.  Figure \ref{fig:angmom}
shows that the angular momentum along the stream is consistent with
being constant within the errors.

\section{Effects of the Galactic Potential}
\label{sec:potential}

We now address the issue of the choice of potential used.  Until now,
we have assumed a simple logarithmic potential giving a constant
circular velocity.  It is instructive to check whether this is an
over-simplification and we should instead be using a potential which
has a power-law dependence on the Galactocentric distance.  As with
the radial velocity correction above, the baseline for our check is
the method given in Section \ref{subsec:masterpoints}.  The following
equation results from using properties of the $H$ point in combination
with a rearrangement of equation (\ref{eqn:energy}) with $\psi(r)
\propto r^{-\alpha}$:

\beqn
1 + \frac{v_H^2 - \dot{r}^2}{v_c^2}\left(\frac{r_H}{\rsun}\right)^\alpha - X^{-2} - \frac{2}{\alpha}\left(1 - X^{-\alpha}\right) = 0\,\,,
\eeqn
where $X = r/r_H$ and all other variables have the same definitions as
before.

When the constant $\alpha$ is set to a small positive quantity, a
small negative correction results for the Galactocentric
distances. For example, $\alpha = 0.01$ shifts the distances found
with the logarithmic potential down by approximately 0.7 kpc and
$\alpha = 0.1$ shifts the distances by approximately 7 kpc.  

It is instructive to find the range of values of $\alpha$ that is
consistent with acceptable values of the Galaxy's mass out to
$100\kpc$, $M_{\rm 100}$.  Imposing the total mass within this radius
and a circular velocity at the Sun automatically imposes the steepness
of the density profile.  Since $\psi(r) \propto r^{-\alpha}$ and $-r^2
\ud\psi/\ud r = G M$, it follows that $G M(r) = v_c^2 \rsun^{\alpha}
r^{1-\alpha}$.  If we ask, for example, that $\alpha$ be consistent
with $0.8 \times 10 ^{12} \Msun< M_{\rm 100} < 2.0 \times 10^{12}
\Msun$, which is consistent with values given by
\cite{1992MNRAS.255..105K}, then $\alpha$ is restricted to values
between 0.14 and $-0.24$.  Employing a smaller mass for the Galaxy
would raise the value of $\alpha$.  If, for example, the mass range
given above were applied at $200\kpc$, then the value of corresponding
$\alpha$ would range between $0.33$ and $0.04$. The actual range is
narrower when physical demands are incorporated, such as that the
Magellanic Clouds should not be at ridiculously small distances and
that we should perhaps expect the circular velocity to taper off
gradually between the solar position and the distance at which the
farthest point on the stream resides. We would, for example, expect
$\alpha$ to be no greater than 0.1 in order that the distance to the
Magellanic Clouds are not too discrepant with observationally
determined values.

\section{Relaxing the Angular Momentum Constraint}
\label{sec:deltah}

In previous sections, we have assumed all of the material comprising
the stream to be following the same orbit, with the same value of
angular momentum and energy being applicable along the entire length
of the stream.  We now investigate the effect of relaxing this
constraint.  If the material forming the Magellanic Stream had been
stripped from the Clouds at the last passage through perigalacticon,
the material must have undergone a change in energy and angular
momentum in order for the observed streaming shape to have formed.
There is an observed lag of approximately $95\deg$ of the end of the
stream relative to the position of the Magellanic Clouds.

In the course of a tidal tearing, the outer parts of the victim are
torn off by the strong tides close to pericentre.  For a small victim,
the tidal debris will have almost the same pericentre as the remnant
of the victim.  Thus the energies and angular momenta of the tidal
debris are related by the condition of all having the same pericentre
distance.  An alternative interesting way of seeing this is to imagine
that near pericentre, where the tearing occurs, it is sufficient to
treat the potential under which the debris moves as that of two bodies
rotating about one another with angular velocity $\Omega_{\rm p}$.  In
reality the rotation rate varies, but it will come to a maximum at
pericentre so it will not be a great error to treat it as constant
near there.  Now an orbit in the field of two bodies rotating around
one another at constant angular velocity $\Omega_{\rm p}$ will have a
constant Jacobi integral (or energy relative to the rotating axes).
The quantity $\varepsilon - \Omega_{\rm p} h$ will be conserved as
debris leave the victim.  This implies that those that gain most
angular momentum $\Delta h$ from lagging just behind the victim will
gain energy $\Omega_{\rm p} \Delta h$.  This argument leads to
essentially the same conclusion as the demand that the perigalactica
of all the debris be the same.  Hence when we deal with debris spread
along the orbit, we change both energy and angular momentum in
proportion and use $\Delta h$ as our parameter.

\subsection{Method}

We continue to use a generalised potential of the following form in
the energy equation, now with an explicit constant:
\beqn
\label{eqn:genpot}
\psi(r) = \frac{K}{r^\alpha}\,.\\
\eeqn
Differentiating the potential and using the fact that the square of
the circular speed at radius $r$ is given by $-r\partial\psi/\partial
r$ gives the equation $v_c^2(r) = \alpha K/r^\alpha$.  Since we have
both a value for the circular speed at the location of the Sun and its
distance from the Galactic centre, $K$ can be calculated given a value
for $\alpha$; the potential at a given radius is then known.  We can
rearrange the energy equation by using $u = 1/r$ to obtain:

\beqn
\label{eqn:u}
u'^2 + u^2 = \frac{2}{h^2}\,\left(\psi(u) + \varepsilon \right)\,,\\
\eeqn
where the prime denotes differentiation with respect to the azimuthal
angle $\phi$.  Using equations (\ref{eqn:energy}), (\ref{eqn:genpot}) and
(\ref{eqn:u}) for the peri- and apogalacticon gives the following
equation for the angular momentum $h$:

\beqn
\label{eqn:h}
h^2 = \frac{2 K (u_{\rm peri}^\alpha - u_{\rm apo}^\alpha)}{u_{\rm peri}^2 - u_{\rm apo}^2}\,\,.\\
\eeqn
Keeping the assumption that the Clouds and stream lie in a planar
configuration, the orbit of the Clouds\footnote{For the purposes of
  this calculation, the Large and Small Magellanic Clouds ---
  collectively referred to as `the Clouds' --- were taken as a
  combined single point at their approximate centre of mass.} can be
computed for a given perigalacticon and apogalacticon.  The constant
$\alpha$ in the potential was set to 0.01.  Note that it is not
strictly necessary to have an absolute value of $K$ if we only require
relative values of angular momenta.

The true current position of the Clouds was estimated to be
approximately 0.7 rad from the perigalacticon that the Clouds have
just passed through; we refer to this angle as the `offset' angle.
Note that more than half of the stream material lies on the other side
of this perigalacticon.  In the orbital computation, we know the time
it has taken for the Clouds to cover the distance from the initial
perigalacticon --- where we assume the material for the stream was
originally stripped --- to their current position.  During the same
period in time, the end of the stream must have covered a smaller
angle in the sky in order to have left the observable lag-angle of
approximately $95\deg$, despite having started at the same location.

\subsection{Results}

We took a fixed perigalacticon distance for the Clouds and
investigated the effect that changing the distance to apogalacticon
has on the total change in angular momentum that occurs along the
stream.  For $r_{\rm apo}$ of between 80 and 180$\kpc$ and a fixed
$r_{\rm peri}$ of 45\kpc, we find a simple relationship between
$r_{\rm apo}$ and $\Delta h/h$ through a fit of the following form:

\beqn
\frac{\Delta h}{h} = \left(\frac{a}{r_{\rm apo}}\right)^b\,;
\,\,a \simeq 19.4\kpc\,;\,b \simeq1.55\,,\\ 
\eeqn
where $\Delta h$ now specifically refers to the absolute change in
angular momentum from the Clouds to a point $95\deg$ downstream.  The
largest difference in angular momentum arises from the least eccentric
orbits.  For an assumed apogalacticon of $120\kpc$ for the Magellanic
Clouds, we find that $\Delta h/h = 0.060$, implying a 6\% difference
in the angular momentum between the extreme ends of the stream.  For
these choices of parameters, the Clouds would be positioned at $r =
51.7\kpc$ and the end of the stream at $59.3\kpc$.  Changing only the
perigalacticon to $44\kpc$ brings the Clouds slightly further in to
$50.6\kpc$ and a slightly smaller value of $\Delta h/h$.  We find that
the distance in an unperturbed orbit to the same azimuthal position
relative to the Magellanic Clouds as the end point is only of order
$0.5\kpc$ less than that for the perturbed orbit.

\subsection{Consistency Checks}

By allowing for a gradient in the angular momentum along the length of
the stream, the angular position of the current perigalacticon will be
different between the Clouds and the end of the stream, even though,
by design, the distance of closest approach to the Galactic centre
will remain the same.  When setting an offset angle for the Clouds,
which is taken relative to the current perigalacticon, it is important
to remember that the location of perigalacticon gradually shifts along
the stream.  We have checked that its difference for the extreme ends
of the stream is merely $1-2\deg$ and hence there is little error
in measuring the offset angle from the perigalacticon of the computed
Magellanic Clouds' orbit, as opposed to an averaged value.

Using the first set of parameters described in the section above, we
have also checked the effect of altering the offset angle by a few
percent to ensure that an error in determining this value from
observational data and calculations described in Section
\ref{subsec:masterpoints} would not lead to a significant error in
$\Delta h/h$.  We find that the percentage difference in this quantity
resulting from a $+10\deg$ change in the offset angle is tiny, on the
order of a few hundredths of a percent.  An under-estimation of the
offset angle by $10\deg$ produces a greater error, on the order of a
few tenths of a percent of $\Delta h/h$.  This is as expected, given
that changes in velocity over the same angle occurs more rapidly when
closer to perigalacticon than when further away. The change in angular
momentum that would be required to keep the end of the stream far
enough away therefore becomes slightly larger.  Even for the
under-estimation case, however, the errors are still relatively small.

One other point to note is that the concept of an $H$ point, defined
previously as the location on the orbital stream with the highest
radial velocity, is now void.  This is because the stream material is
not following the same orbit as that of the Clouds.  A check to see
whether the projected line-of-sight velocity at the tip of the stream,
earlier coined the $H$ point, is consistent with the observed value at
that location served as an independent test of the validity of the
approach used to exploit the lag-angle in calculating the angular
momentum deviation.

\section{Summary and Conclusions}
\label{sec:sum}

We have applied a variety of geometrodynamical methods to streams of
high-velocity clouds in an attempt to constrain their distances.  Some
of these methods, including the near method, have the nice feature
that distance determinations would be possible without the parameter
dependency inherent in any simulations.  We find, however, that the
near method in particular becomes rather inaccurate when the angle
subtended at the stream by $\vecrsun$ becomes small.  At the expense
of becoming more model dependent by assuming a halo form for the
Galactic potential, we have improved on this accuracy.  We then find
distances to the entire length of the Magellanic Stream by assuming
that the gas comprising the stream nearly follows a planar orbit about
the Galactic centre.  We find the perigalacticon to be at $45\kpc$ and
the maximum distance at the trailing end to be $75\kpc$.  The
heliocentric distances are not very different from the Galactocentric
distances, varying typically by $1\kpc$ or less.  The centre-of-mass
of the Magellanic Clouds would be at $47\pm5\kpc$ from us, according
to this method; this is consistent with various observationally
determined values for the Large Magellanic Cloud (see
\citet{2004NewAR..48..659A} for a review).  Application of our methods
to Complex A and C did not yield very promising results under the
assumptions that were made.  This indicates that our hypothesis,
namely that the streams were nearly following an orbit around the
Galactic centre in a well-defined plane, was not valid for these two
streams of high-velocity clouds.

In this paper, we have tested the idea that streams of HVCs follow
orbits.  However the assumption that the orbit lies in a plane that
passes through the Galactic centre is rather restrictive when applied
to nearby streams that do not pass close to the Galactic Pole. It is
possible to remove the restriction of a planar orbit, provided a model
of the Galaxy's potential is given. We find that the transverse
velocity of the stream can be deduced from the run of radial
velocities as a function of angle along the stream, provided the
distance to one point of the stream is assumed. Orbits can then be
computed in the given potential. Both directions in the sky and radial
velocities are then computed and these are compared with the
observations.  Such a method and an example of its application to
Complex A and the stellar Orphan stream has already been presented
\citep{2007MNRAS.378L..64J}.  We note that such methods of deriving
the transverse velocity component were used in a related problem by
\cite{1977A&A....57..265F}, but probably date back to an earlier
century.  We are also exploring the possible existence of stream
associations.  Following on from these works, we hope, eventually, to
give a definitive answer to the question of whether high-velocity
cloud streams follow orbits around the Galaxy.

\subsection*{Acknowledgements}
The authors thank B. Wakker and R.~D.~Davies for the kind use of their
data for Complex A, C and the Magellanic Stream and the anonymous
referee for helpful comments.  SJ acknowledges financial support from
PPARC.

\appendix
\section{Example of the Near Method}
\label{sec:nearmethodex}

We have tested the near method (see Section \ref{subsec:near}) on
simulated orbits of varying orientations and distances in the
logarithmic potential.  The constraint on $\vartheta_d$ given in
Section \ref{subsec:near} was deduced from these test cases; one
working example is shown here.  The normal to the orbital plane is
given by $\normn = (-0.687, 0.725, 0.047)$.  The orbit is nearly polar
but is otherwise inclined at approximately $43\deg$ to the Galactic
$x$ axis.

\begin{figure}
\includegraphics[width=0.45\textwidth]{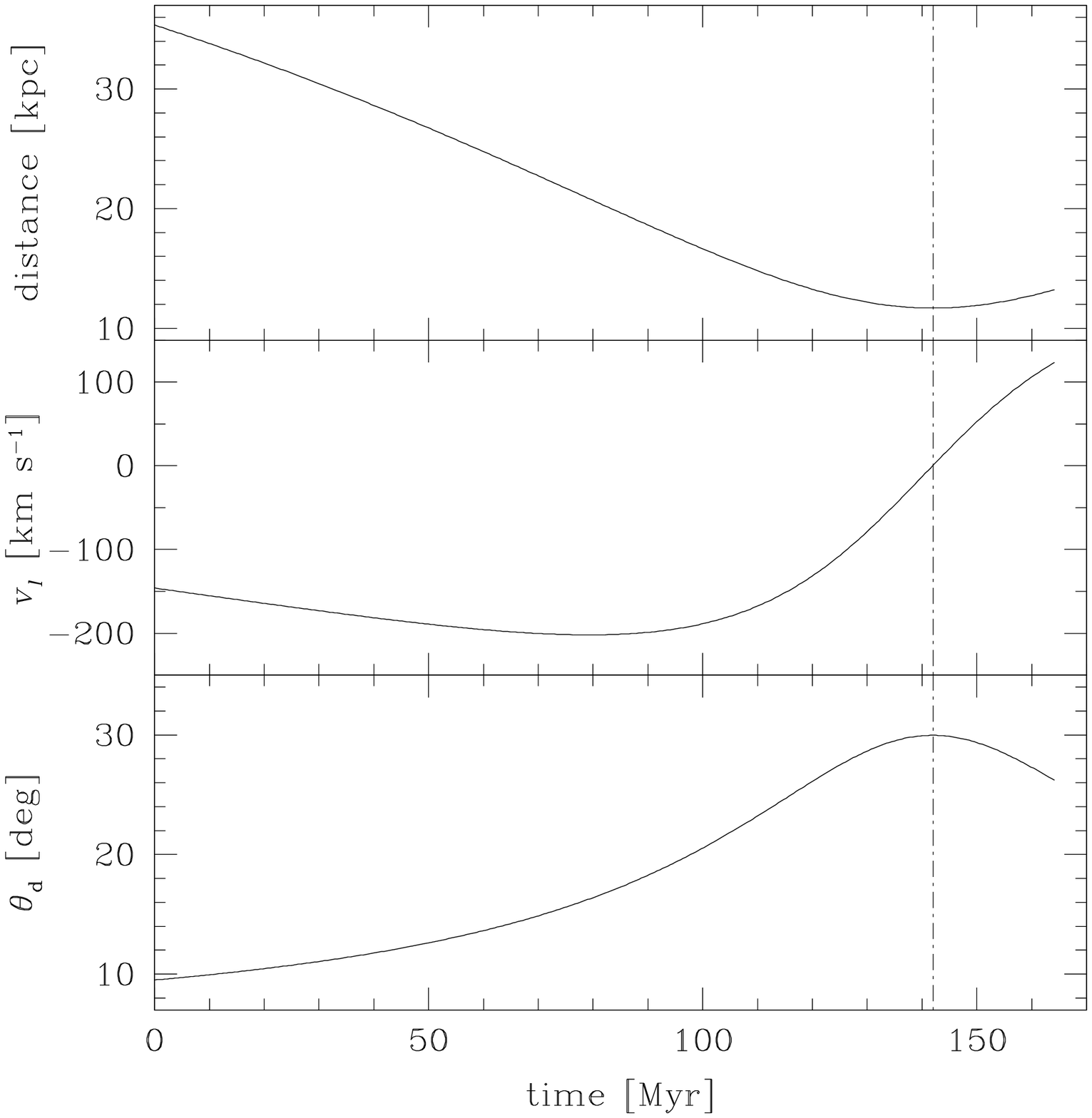}
\caption{Top: heliocentric distance to the orbital points over time.
  Middle: line-of-sight velocity along the orbit.  Bottom: the angle
  $\vartheta_d$ subtended at the stream by $\vecrsun$ projected onto
  $\normn$.  Initial conditions: Galactocentric spherical polar
  coordinates of $r = 31\kpc$ and $(\theta,\phi) = (143,47)\deg$;
  ${\bf{v}} = (-151,-143,0.2)\kms$. The time at which the $Z$ point is
  reached is denoted by the dot-dash line.  }
\label{fig:nearmethodex}
\end{figure}

\begin{figure}
\includegraphics[width=0.45\textwidth]{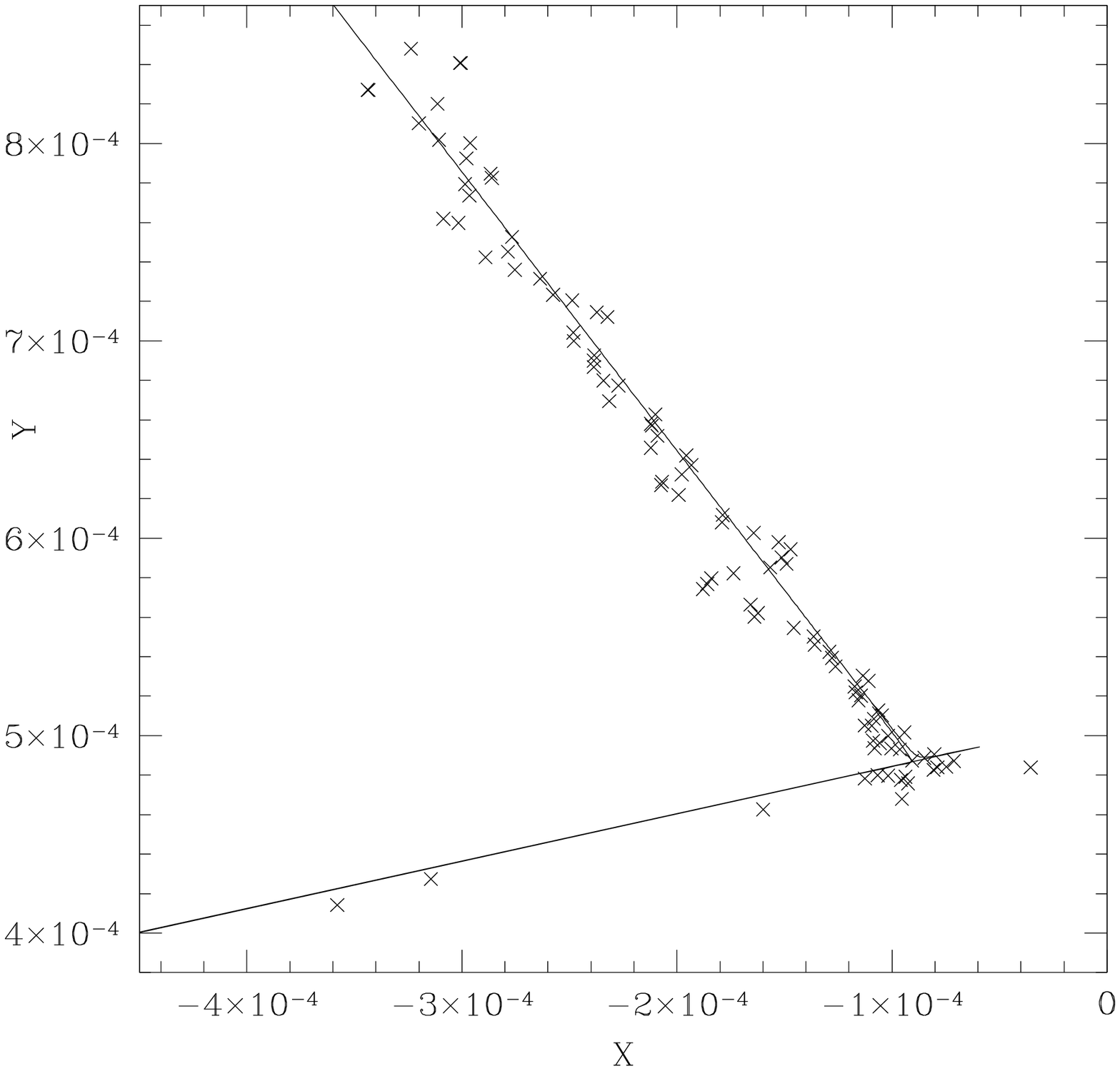}
\caption{The variables $X$ and $Y$ of equation (\ref{eqn:XY}) have
  been plotted against one another for the orbit shown in figure
  \ref{fig:nearmethodex}.  The straight solid lines show the results
  from an exact orbit (where, for clarity, individual points are not
  shown).  The crosses give the results for the same orbit, but where
  the direction cosines have had a random scatter (of maximum 0.01)
  applied, before the orbit is smoothed by the code.  The few points
  along the lower solid line indicate the behaviour of the $X$ and $Y$
  variables for positions near the $Z$ point.}
\label{fig:XYplot}
\end{figure}

Figure \ref{fig:nearmethodex} shows how the heliocentric distance to
the orbit, the line-of-sight velocity and the angle $\vartheta_d
=\mathrm{arcsin}(\vecrsun.\normn/d)$ change as a function of time.
The orbit shown has a length of $130\deg$.  The singular behaviour of
the variables $X$ and $Y$ near the $Z$ point is clearly visible in
Figure \ref{fig:XYplot}, both when using the exact orbital data and
when some scatter has been applied to the direction cosines.  The
distance to the $Z$ point, $d_Z$, can be calculated from the noisier
plot to within a few per cent of the correct value.

\end{document}